\documentclass[10pt,conference]{IEEEtran}
\usepackage{spconf,amsmath,graphicx}
\usepackage[utf8]{inputenc}
\usepackage{glossaries}                
\usepackage{hyperref}
\hypersetup{
    colorlinks=true,
    linkcolor=blue,
    filecolor=magenta,      
    urlcolor=cyan,
}
\newcommand{\thing}{FlashLight CNN}
\newcommand\figureSize{.16}
\newcommand{\psnr}{PSNR}
\newcommand{\ssim}{SSIM}

\usepackage{tikz}
\usetikzlibrary{shapes}
\usetikzlibrary{arrows,decorations.pathreplacing,decorations.pathmorphing,backgrounds,positioning,fit,petri,chains}
\usetikzlibrary{spy}
\usetikzlibrary{calc}
\usetikzlibrary{trees}
\usetikzlibrary{fit}
\usepackage{float}
\usepackage{ifthen}
\usepackage{pgfplotstable}
\usepackage{booktabs}
\usepackage{graphicx}
\usepackage{multirow}
\pgfplotstableset{
every head row/.style={ 
before row=\toprule, %
after row=\midrule %
}, %
every last row/.style={ %
after row=\bottomrule %
}, %
fixed, 
fixed zerofill, %
precision=2, %
}
\def\showdetail#1#2#3#4#5#6#7#8#9{%
\node[image,#2] (#1) {\includegraphics[width=#6]{#1}};
\path (#1.south west) ++(#3) coordinate (#1 detail position);
\path (#1.south west) ++(#7) coordinate (#1 second detail position);
\path (#1.south west) ++(#8) coordinate (#1 third detail position);
\spy[red] on (#1 detail position) in node [above=of #1] (#1 first spy position);
\node[below=1mm of #1] (title) {#4\hspace{0pt}#9};
}
\def\printpsnr{~\pgfmathprintnumber{\pgfplotsretval}~dB}
\def\showdetailall#1#2#3#4#5#6#7#8#9{%
\def\imw{#3} 
\begin{tikzpicture}[
spy using outlines={%
rectangle,
magnification = #6,
width = #4,
height = #5,
very thick,
},
node distance = #7,
image/.style = {anchor=south west, inner sep=0},
]

\pgfplotstableread{#1_sigma.csv}\psnrtable
\pgfplotstablegetelem{0}{noisy}\of\psnrtable
\showdetail{#1}{}{#2}{Ground truth}{#5}{#3}{#8}{#9}{}
\showdetail{#1_z}{right = of #1}{#2}{Noisy}{#5}{#3}{#8}{#9}{\printpsnr}
\pgfplotstablegetelem{1}{noisy}\of\psnrtable
\showdetail{#1_bm3d}{right = of #1_z }{#2}{BM3D}{#5}{#3}{#8}{#9}{\printpsnr}
\pgfplotstablegetelem{2}{noisy}\of\psnrtable
\showdetail{#1_dncnn}{right = of #1_bm3d}{#2}{DnCNN}{#5}{#3}{#8}{#9}{\printpsnr}
\pgfplotstablegetelem{3}{noisy}\of\psnrtable
\showdetail{#1_flashlight}{right = of #1_dncnn}{#2}{FlashLight\hspace{-4pt}}{#5}{#3}{#8}{#9}{ \printpsnr}
\end{tikzpicture}
}

\newacronym{awgn}{AWGN}{additive white Gaussian noise}
\newacronym{bm}{BM}{Block matching}
\newacronym{bm3d}{BM3D}{Block Matching 3D}
\newacronym{bmcnn}{BMCNN}{block matching convolutional neural network}
\newacronym{cnn}{CNN}{convolutional neural network}
\newacronym{cnnf}{CNNF}{\gls{cnn}-based filter}
\newacronym{gpu}{GPU}{graphics processing unit}
\newacronym{gpus}{GPUs}{Graphics Processing Units}
\newacronym{mse}{MSE}{mean squared error}
\newacronym{nlf}{NLF}{nonlocal filter}
\newacronym{nn}{NN}{neural networks}
\newacronym{nlm}{NLM}{Nonlocal Means}
\newacronym{wsd}{WSD}{Wiener filter in Similarity Domain}
\newacronym{psnr}{PSNR}{peak signal to noise ratio}
\newacronym{ssim}{SSIM}{structural similarity index measure}
\newacronym{dcnn}{DCNN}{Deep convolutional neural networks}
\newacronym{dncnn}{DnCNN}{Denoising convolutional neural network}
\newacronym{gan}{GAN}{Generative Adversarial Networks}
\newacronym{bn}{BN}{Batch Normalization}
\newacronym{relu}{ReLu}{Rectified linear unit}
\newacronym{dnn}{DNN}{Deep Neural Networks}
\newacronym{ml}{ML}{Machine Learning}

\usepackage[normalem]{ulem} 
\usepackage{xcolor}         

\newcommand\kacolor{\textcolor[rgb]{0.12,0.74,0.36}}
\newcommand\alcolor{\textcolor[rgb]{.6,0.5,0}}
\newcommand\cccolor{\textcolor{orange}}
\newcommand\bicolor{\textcolor{blue}}

\newcommand\kaWuline{\bgroup\markoverwith{\kacolor{\rule[-0.4ex]{2pt}{0.4pt}}}\ULon}
\newcommand\alWuline{\bgroup\markoverwith{\alcolor{\rule[-0.4ex]{2pt}{0.7pt}}}\ULon}
\newcommand\ccWuline{\bgroup\markoverwith{\cccolor{\rule[-0.5ex]{2pt}{0.4pt}}}\ULon}
\newcommand\biWuline{\bgroup\markoverwith{\bicolor{\rule[-0.5ex]{2pt}{0.4pt}}}\ULon}

\title{FLASHLIGHT CNN IMAGE DENOISING}

\name{Pham Huu Thanh Binh $^1$, Cristóvão~Cruz $^1$, Karen Egiazarian ${^{1,2}}$
}
\address{
$^1$ Noiseless Imaging Ltd, Tampere, Finland, \\
$^2$ Faculty of Information Technology and Communication Sciences, Tampere University, Finland
\\
}

\begin{document}
\maketitle
\begin{abstract}
This paper proposes a learning-based denoising method called FlashLight CNN (FLCNN) that implements a deep neural network for image denoising. The proposed approach is based on deep residual networks and
inception networks and it is able to leverage many more parameters than residual networks alone for denoising grayscale images corrupted by \gls{awgn}. 
FlashLight CNN  demonstrates state of the art performance when compared quantitatively and visually 
 with the current state of the art image denoising methods.

\end{abstract}

\section{Introduction}
\label{sec:intro}
Image denoising is a fundamental problem in image processing aiming at reconstructing an image from its noisy measurement. Since 2005 for a decade this field has been dominated by non-local transform domain patch based methods, such as BM3D \cite{dabov_2007_image}, \cite{rus_2005}, and its modifications,  BM3D-SAPCA \cite{BM3D-SAPCA} and WNNM \cite{gu_2014_weighted}. Most notably, BM3D has been the state of the art method until the recent rapid advancement of \gls{ml} based approaches, more specifically, \gls{dnn} based approaches. In contrast with the \emph{traditional} approaches, \gls{dnn} based solutions employ a vast dataset of examples and learn how to invert the degradation function \cite{jain_2009_natural}. These methods saw initial success in the field of computer vision~\cite{he_2016_deep,alexnet,kim_2016_accurate,goodfellow_2014_generative}, but it was quickly realized that they could also be used in image denoising and other image restoration tasks. While in the field of computer vision the network learns a mapping between input image and image label \cite{alexnet}, 
when applied as a denoiser it instead learns a mapping between a degraded and a clean image \cite{zhang_2017_gaussian}. 
The introduction of techniques such as residual learning \cite{he_2016_deep} and batch normalization \cite{ioffe_2015_batch} allowed the depth of these networks to increase over time along with their performance, also in the field of image denoising \cite{zhang_2017_gaussian}. Currently, the state of the art in image denoising is dominated by \gls{dncnn} based methods \cite{zhang_2017_gaussian} and its modifications, FFDNet~\cite{zhang_2017_ffdnet}, IRCNN~\cite{zhang_2017_learninga}, HRLNet ~\cite{hrlnet} and others.
Despite their recent success, \gls{dnn} based approaches suffer from diminishing feature reuse and are unable to take advantage of an increased number of parameters, be it either by increasing the number of layers or using wider kernels per layer. Furthermore, they have been shown to exhibit a narrow receptive field \cite{luo_2016_understanding} which limits their ability to take advantage of long range correlations.


This paper proposes a \gls{cnn} a network called FlashLight CNN inspired by \gls{dncnn}~\cite{zhang_2017_gaussian} and Inception-ResNet~\cite{szegedy_2017_inception} architectures that solves the above-mentioned issues. The main goal of the proposed network is to overcome the dimishing feature reuse by use of inception layers in such a way that an increase in the number of parameters of the network leads to increased performance. Additionally, by using layers with much wider support, we are able to effectively increase the receptive field of the network and its ability to restore image content.
%
%
The proposed approach demonstrates state of the art performance when compared to current image denoising methods.

\begin{figure*}[t]
\centering
\begin{tikzpicture}
[
    rect1/.style={rectangle,draw=black!70,thick,inner sep=2pt,minimum size=1.5cm},
    circ/.style={circle,draw=black!70,inner sep=2pt,minimum size=0.7cm},
    node distance=0.8cm,
    thick,
    font=\bfseries,
    brace/.style={decorate,decoration={brace,amplitude=3pt,mirror},rotate=180},
]
\node[] (input_x)[]{};

\node[rect1] (step1)[right=of input_x]{};
\coordinate[right=1cm of input_x] (cont);
\node[rect1] (step2) [right=of step1, label=below:{l layers}] {};
\node[rect1] (step3) [right=of step2, label=below:{m layers}] {};
\node[rect1] (step4) [right=of step3, label=below:{n layers}] {};

\node[rect1] (step5) [right=of step4] {\rotatebox{90}{\parbox{4cm}{}}};
\node[circ] (plus) [right=0.5cm of step5] {+};


\node[align=center,font=\large] at (step1.center) {conv\\$3 \times 3$\\$64$};
\node[align=center,font=\large] at (step2.center) {conv\\$3 \times 3$\\$64$};
\node[align=center,font=\large] at (step3.center) {conv\\$5 \times 5$\\$64$};
\node[align=center,font=\large] at (step4.center) {incept\\A};
\node[align=center,font=\large] at (step5.center) {conv\\$3 \times 3$\\$1$};

\draw [->] (input_x)--(step1);
\draw [->] (step1.east)--(step2.west);
\draw [->] (step2.east)--(step3.west);
\draw [->] (step3.east)--(step4.west);
\draw [->] (step4.east)--(step5.west);
\node[] (output) [right=of plus] {};


\begin{scope}[on background layer]
    \coordinate[above left=0.2cm of step2] (a);
    \coordinate[above right=0.2cm of step3] (b);
    \coordinate[below right=0.2cm of step3] (c);
    \coordinate[below left=0.2cm of step2] (d);
    \fill[orange!30] (a) -- (b) -- (c) -- ++(0,-0.4) -| (d);
\end{scope}

\draw[decorate,decoration={brace,amplitude=3pt,mirror},rotate=180] 
    (-6.8,-1.1) coordinate (t_k_unten) -- (-3.4,-1.1) coordinate (t_k_opt_unten); 
\node at (5.2,1.6){Warm-up phase};
\draw[decorate,decoration={brace,amplitude=3pt,mirror},rotate=180] 
    (-9.6,-1.1) coordinate (t_k_unten) -- (-7.7,-1.1) coordinate (t_k_opt_unten); 
\node at (8.6,1.6){Boost phase};
\begin{scope}[on background layer]
    \coordinate[above left=0.2cm of step4] (a);
    \coordinate[above right=0.2cm of step4] (b);
    \coordinate[below right=0.2cm of step4] (c);
    \coordinate[below left=0.2cm of step4] (d);
    \fill[blue!30] (a) -- (b) -- (c) -- ++(0,-0.4) -| (d);
\end{scope}

\draw [->] (step5) -- (plus);
\draw [->]  (step1.west) ++(-0.3,0) -- ++(0,-1.5) -| (plus);
\draw [->] (plus) -- (output);

\draw (input_x.east) ++(0.25,0) node[above] {$z$};
\draw (plus.west) ++(-0.25,0.25) node {$-$};
\draw (plus.south) ++(-0.25,-0.25) node {$+$};
\draw (plus.east) ++(0.25,0) node[above] {$\hat{y}$};

\end{tikzpicture}
\caption[The proposed Flash Light CNN architecture]{The proposed Flash Light CNN architecture for denoising, with noisy input $z$ and estimate $\hat{y}$. The orange background marks the \emph{warmup} phase while the blue background marks the \emph{boost} phase. Batch normalization and \emph{relu} units were omitted for sake of clarity.}
\label{fig:flashflightCNN}
\end{figure*}
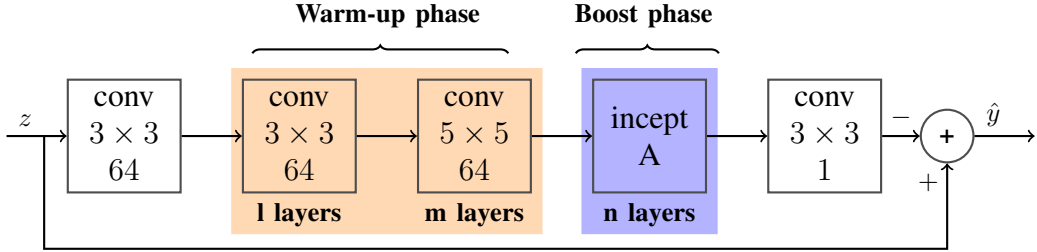

\section{Background}

\label{sec:inception}

With the increased availability of computational resources, \glspl{cnn} have the opportunity to grow and employ more and more parameters \cite{szegedy_2015_going}. However, naïve approaches to increase the number of parameters of a \glspl{cnn} by increasing the number of layers has resulted in decreased performance. This effect has been blamed in issues such as diminishing feature reuse and narrow receptive fields \cite{he_2016_deep}. Several solutions have been proposed for these issues, which include the use of skip connections \cite{he_2016_identity} and wide-residual layers \cite{he_2016_deep}.



Skip connections allow a network to learn a, so called, residual mapping.  They consist of an identity mapping placed between two non-adjacent layers\cite{he_2016_identity}. \cite{he_2016_deep} showed that when these connections are used in every layer, increasing the network depth translates into performance gains, as opposed to performance loss observed when these connections are not used. We also know from \gls{dncnn}~\cite{zhang_2017_gaussian} that even when considering shallower networks, using just one skip connection between the input and the output of the whole network improves performance. These networks are called residual networks.

Increasing the depth of the networks also
slows down training and can lead to diminishing feature reuse \cite{zagoruyko_2016_wide}. So on top of using skip connections,  \cite{zagoruyko_2016_wide} propose that layers should also be made wider and thicker by adding more feature planes and more convolution kernels per network layer, so each network layer would contain many more parameters, leading to the wide residual layers. Networks equipped with these wide residual layers performed better than networks with the \emph{regular} layers, when the number of parameters remained constant. An added bonus was that these shallower networks train much faster. Bottom line, if one can afford more parameters, one should not add more layers, but make them wider and thicker instead.

One notable network that successfully combined several of these techniques is the Inception Network \cite{szegedy_2017_inception}. The Inception Network has gained in popularity since 2014 when it achieved the top position in the ILSVRC 2014 competition~\cite{russakovsky_2015_imagenet}. There are different versions of it, with the latest ones being Inception-v4 and Inception-Resnet~\cite{szegedy_2017_inception}. On top of the use of skip connections and wide-residual layers, Inception-Resnet also uses cascades of small kernels as opposed to big kernels in an attempt to reduce the overall number of parameters while maintaining the depth of the networks~\cite{szegedy_2015_going}.


\section{Proposed method: Flashlight CNN}
\label{sec:flashlight}
We propose \thing, a network architecture that combines elements from DnCNN and Inception-Resnet to combat diminishing feature reuse and successfully leverage a significantly increased number of parameters for the task of denoising grayscale images corrupted by \gls{awgn}.
\begin{figure}[!ht]
\begin{center}
\begin{tikzpicture}
[
common/.style={draw=black!70,fill=white!100,thick,text centered,text width=1cm},
rect1/.style={common,rectangle,},
rect2/.style={rect1},
circ/.style={common,circle},
node distance=0.4cm, thick,scale=1, every node/.style={scale=0.7},
font=\bfseries]

\coordinate (input);
\node(input_previous) [above=of input] {Input from the previous layer};




\node[rect1] (branch2_1) [below left=of input] {{1x1 \\32}};
\node[rect1] (branch2_2) [below=of branch2_1] {{3x3 \\48}};
\node[rect1] (branch2_3) [below=of branch2_2] {{3x3 \\64}};

\node[rect2] (branch1_0) [left=of branch2_1] {{1x1 \\32}};
\node[rect2] (branch1_1) [below=of branch1_0]          {{3x3 \\32}};
\node[rect2] (branch1_2) [below=of branch1_1]          {{3x3 \\48}};
\node[rect2] (branch1_3) [below=of branch1_2]          {{3x3 \\64}};
\node[rect2] (branch1_4) [below=of branch1_3]          {{3x3 \\64}};

\node[rect1] (branch3_1) [right=of branch2_1] {1x1 \\32};
\node[rect1] (branch3_2) [below=of branch3_1] {3x3 \\64};

\node[rect1] (branch4_1) [right=of branch3_1] {1x1 \\32};

\node[rect1] (concat) [below = 2.5 cm of branch2_3,xshift=1cm,text width=6cm] {{1x1 \\64}};

\node[circ] (plus) [below=of concat] {+};
\node[rect1] (relu) [below=of plus] {relu};
\node[] (output) [below=of relu] {Output to the next layer};

\draw [->] (input.west) -- ++(3,0) |- (plus);

\draw (input_previous) --  (input);
\draw [->] (input) -|  (branch1_0);
\draw [->] (input) -|  (branch2_1);
\draw [->] (input) -|  (branch3_1);
\draw [->] (input) -|  (branch4_1);

\draw [->] (branch1_0) -- (branch1_1);
\draw [->] (branch1_1) -- (branch1_2);
\draw [->] (branch1_2) -- (branch1_3);
\draw [->] (branch1_3) -- (branch1_4);
\draw [->] (branch1_4) -- (branch1_4 |- concat.north);
\draw [->] (branch2_1) -- (branch2_2);
\draw [->] (branch2_2) -- (branch2_3);
\draw [->] (branch2_3) -- (branch2_3 |-concat.north);
\draw [->] (branch3_1) -- (branch3_2);
\draw [->] (branch3_2) -- (branch3_2|- concat.north);
\draw [->] (branch4_1) -- (branch4_1 |- concat.north);

\draw [->] (concat) -- (plus);
\draw [->] (plus) -- (relu);
\draw [->] (relu) -- (output);

\end{tikzpicture}
\end{center}
\caption{Inception layer used in the proposed architecture.}
\label{fig:inception}
\end{figure}
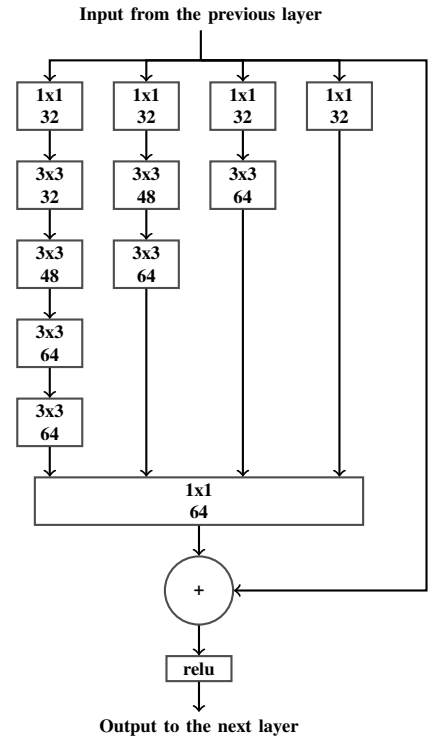

\begin{table*}[!t]
\begin{center}
\centering
\pgfplotstableread{data_csv/set12_performance_ssim.csv}\psnrsettwelve
\pgfplotstableread{data_csv/bsd68_performance_ssim.csv}\psnrbsdsixtyeight
\pgfplotstableread{data_csv/urban100_performance_ssim.csv}\psnrurbanhundred
\pgfplotstablevertcat{\fulltable}{\psnrsettwelve}
\pgfplotstablevertcat{\fulltable}{\psnrbsdsixtyeight}
\pgfplotstablevertcat{\fulltable}{\psnrurbanhundred}
\pgfplotstabletypeset[
	create on use/dataset/.style={create col/set list={Set12,,,BSD68,,,Urban100,,}},
	columns/dataset/.style={
		string type,
		column name=Dataset,
		assign cell content/.code={
			\pgfmathparse{int(mod(\pgfplotstablerow, 3))}%
			\ifnum\pgfmathresult=0%
				\pgfkeyssetvalue{/pgfplots/table/@cell content}
				{\multirow{3}{*}{\emph{##1}}}%
			\else
				\pgfkeyssetvalue{/pgfplots/table/@cell content}{}%
			\fi
		},
	},
	columns/sigma/.style={column name=$\sigma$},
	columns/bm3d/.style={column name=\psnr},
    columns/bm3dssim/.style={column name=\ssim,precision=4},
	columns/dncnn/.style={column name= PSNR},
    columns/dncnnssim/.style={column name= SSIM,precision=4},
	columns/ffdnet/.style={column name=\psnr},
	columns/ffdnetssim/.style={column name=\ssim,precision=4},
	columns/hrlnet/.style={column name=\psnr},
	columns/hrlnetssim/.style={column name=\ssim,precision=4},
	columns/ircnn/.style={column name=\psnr},
	columns/ircnnssim/.style={column name=\ssim,precision=4},
	columns/flashlightcnn/.style={column name=\psnr},
	columns/flashlightcnnssim/.style={column name=\ssim,precision=4},
    every head row/.style={
        before row=\toprule
        & \multicolumn{1}{c}{}
        & \multicolumn{2}{c}{BM3D~\cite{dabov_2007_image}}
        & \multicolumn{2}{c}{DnCNN~\cite{zhang_2017_gaussian}}
        & \multicolumn{2}{c}{FFDnet~\cite{zhang_2017_ffdnet}}
        & \multicolumn{2}{c}{IRCNN~\cite{zhang_2017_learninga}}
        & \multicolumn{2}{c}{HRLNet~\cite{hrlnet}}
        & \multicolumn{2}{c}{FLCNN}
         \\
        ,after row=\midrule,
    },
	every nth row={3}{before row=\midrule},
    columns/sigma/.append style={precision=0},
	columns/hrlnet/.append style={string replace={0}{}, empty cells with={---}},
	columns/hrlnetssim/.append style={string replace={0}{}, empty cells with={---}},
	columns={dataset, sigma, bm3d,bm3dssim, dncnn,dncnnssim, ffdnet,ffdnetssim, ircnn,ircnnssim, hrlnet, hrlnetssim,flashlightcnn,flashlightcnnssim}
]
{\fulltable}

\caption{Performance comparison in terms of \gls{psnr} and \gls{ssim} on Set12, BSD68 and Urban100 with noise levels of 15, 25, 50. The unavailable values are replaced by "---".} 
\label{table:data_denoise}
\end{center}
\end{table*}

\thing\ is made up two phases: \emph{warmup} and \emph{boost}, with a residual skip connection between the input and the output, as shown in Fig.~\ref{fig:flashflightCNN}. The \emph{warmup} phase uses only \emph{conventional} convolutional layers and resembles a \emph{typical} \gls{cnn}. The \emph{boost} phase on the other hand, uses much wider residual inception layers that rapidly increase the number of parameters of the network  while avoiding the diminishing feature reuse that would ensue if only conventional convolutional layers would be employed. The inception layers used in this network, shown on Fig.~\ref{fig:inception}, were based on the work of \cite{szegedy_2017_inception}. They employ input dimensionality reduction as a way to reduce the computational complexity. They also use cascades of smaller filter banks instead of a single big filter in order to reduce the number of parameters required by each layer. Finally, they sport a residual skip connection that has been shown to be an effective way of avoiding the diminishing feature reuse that comes with the increase of the number of parameters in the network.

\begin{figure}[!b]
\centering
\begin{tikzpicture}[scale=0.76]
\begin{axis}[
    title=Validation performance vs number of parameters,
    xlabel={Number of parameters},
    ylabel={PSNR},
    grid=both,
    legend style={at={(0.43,0.1)},anchor=east}
    ]

\pgfplotstablesort[sort key=Params, sort cmp=int >]{\datatablesorted}{data_csv/inception_cnn_models_div2k.csv}
\addplot [color = blue, only marks,mark=square*]  table [x=Params,y=PSRN]
{\datatablesorted};
\addlegendentry{FlashLight CNN} 

\addplot[color= red,   only marks,mark=pentagon ] table[x=Params,y=PSRN]
{data_csv/inception_cnn_models_dncnn.csv};
\addlegendentry{DnCNN like}

\end{axis}
\end{tikzpicture}
\caption{Validation performances vs number of parameters, when the number of parameters of \gls{nn} increases from one to about two million parameters. The performance of \emph{DnCNN} like network decreases drastically, while FlashLight CNN sees increased performance.}
\label{fig:performance_vs_parameters}
\end{figure}
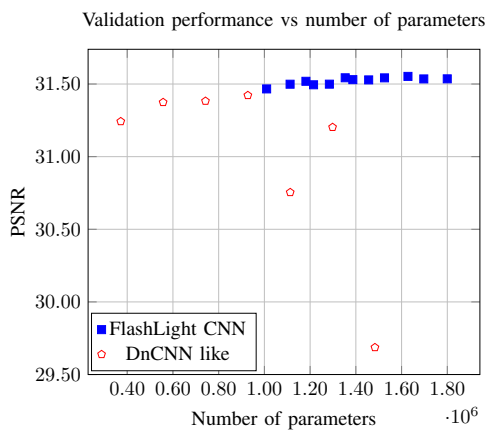

The \emph{warmup} phase is composed of two stages of layers, with the first stage employing only $3 \times 3$ kernels and the second stage employing bigger $5 \times 5$ kernels. Both of the stages extract $64$ features per layer. The purpose of this phase is to extract low level features from the input image that can then be processed by the much more capable \emph{boost} phase. The use of wider kernels in the second stage allows the gradual widening of the receptive field before the boost phase. The \emph{boost} phase in turn uses residual \emph{inception} layers, the model of which is shown in Fig.~\ref{fig:inception}. The combination of all these features leads to a network that uses more processing as the level of abstraction increases, that is, as we move away from the pixel domain in the input. The network is also progressively wider allowing longer range connections to be established as the abstraction level increases towards the output. The progressive widening of the layer's receptive field inspired us to name the network: \thing.

After having defined the overall architecture of the network, the only remaining free parameters are the number of layers in each stage, identified in Fig.~\ref{fig:flashflightCNN} by $l, m, n$. We used exhaustive search to find the best set of parameters, but in order to keep this search tractable, we set constraints on the values taken by each parameter based on our expectation of the network behaviour. We fixed $l = 5$ and set the search space for the other parameters as $m \in \{3,4,5\}, n \in \{3,4,5,6,7\}$.

During the search for the best architecture, we observed that an increased number of parameters translated, up to a certain point, to increased performance, as can be observed in Fig.~\ref{fig:performance_vs_parameters}. This behaviour confirms that the proposed architecture is indeed able to leverage the extra parameters that are made available to the network. Furthermore, we also infer from the comparison with a \emph{DnCNN} like network, which corresponds roughly to our \emph{warmup} phase, that the \emph{boost} phase is essential to the increased performance.


Based on the validation performance of these experiments presented in Fig.~\ref{fig:performance_vs_parameters}, our final configuration is defined by: $l = 5, m = 4, n = 6$, to a total of 15 layers and 1627905 trainable parameters.

During the training process we used the \gls{mse} loss function, 55 epochs, epoch length 4096 and batch size 64. The network weights are initialized by orthogonal method [57] and we use the Adam optimizer. We set the initial learning rate to $1 \times 10^{-3}$ and modulate the learning rate using a step function that drops to $1 \times 10^{-4}$ after 30 epochs. We use batch normalization \cite{ioffe_2015_batch} before every \gls{relu} activation function with exception of the first and last layers. We trained and validated using the DIV2K dataset training and validation splits respectively~\cite{Timofte_2017_CVPR_Workshops}.

\begin{figure*}[]
    \centering
    \showdetailall{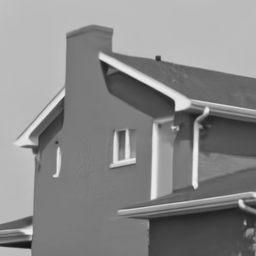}{1.413cm,1.689cm}{\figureSize\textwidth}{\figureSize\textwidth}{1.3cm}{8}{1mm}{1.28cm,2.20cm}{1.623cm,1.40cm}
    \showdetailall{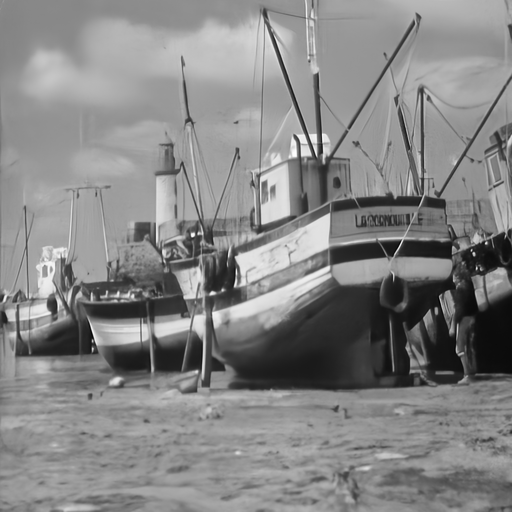}{1.867cm,1.564cm}{\figureSize\textwidth}{\figureSize\textwidth}{1cm}{8}{1mm}{1.28cm,2.80cm}{1.623cm,1.40cm}
    \showdetailall{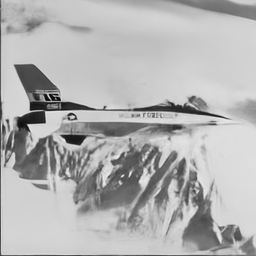}{1.689cm,1.573cm}{\figureSize\textwidth}{\figureSize\textwidth}{1cm}{9}{1mm}{1.28cm,2.80cm}{1.623cm,1.40cm}
    \caption{Visual results with corresponding \gls{psnr} for the house and boat with $\sigma\!=\!25$
and the plane with $\sigma\!=\!50$
on \emph{Set12}.
}
\label{fig:visual_results}
\end{figure*}

\section{Experimental evaluation}
\label{sec:results}
Our introduced FlashLight CNN is evaluated over three common datasets: Set12, BSD68 and Urban100
with \gls{awgn} noise levels of $\sigma \in {15, 25, 50}$ and compared with state-of-the-art methods, namely, BM3D \cite{dabov_2006_image} , DnCNN \cite{zhang_2017_gaussian},  FFDNet~\cite{zhang_2017_ffdnet}, IRCNN~\cite{zhang_2017_learninga}, and HRLNet~\cite{hrlnet}.
The results of evaluation in terms of 
\gls{psnr} and \gls{ssim} metric values are depicted in Table \ref{table:data_denoise}. Our proposed method exhibits better performance than  the other methods in the comparison. Notably it performs significantly better than DnCNN 
for all noise levels and datasets.


Fig.~\ref{fig:visual_results} shows several examples to demonstrate the visual performances of the proposed solution.
The proposed method recovers better the edge patterns than other competing methods. In the \emph{house} and \emph{boat} images, the line shadow and the line below the text are more effectively recovered.
For the stronger noise level $50$ used in the \emph{plane} our proposal exhibits superior performance recovering the text.


The code and models used for this evaluation can be downloaded in \href{https://github.com/binhpht/flashlightCNN}{https://github.com/binhpht/flashlightCNN}.
\section{Discussion and conclusion}
\label{sec:discussion}
We presented \thing, a deep neural network that is able to leverage more parameters than residual networks for the task of image denoising. We showed that the performance of the proposed network increases as the number of parameters is increased. However, the increased number of parameters come with an increased computational cost. While a \gls{dncnn} like network with 557057 parameters takes 1.55 seconds to process all images in \emph{Set12}, \thing, with 1361649 parameters, takes 4.48 seconds, on an NVIDIA GeForce GTX 1080 Ti. Furthermore, experimental results showed that the performance of \thing\ stops increasing after roughly $1.6$ million parameters. It would be worth investigating how to overcome this barrier. Finally, the proposed solution has the potential to be successfully applied to other image processing tasks, such as multispectral image denoising, image super-resolution or deblurring, with minimal modifications. 
\section{Acknowledgements}
This work is in part supported by the Business Finland (project 3418 - E!7632 ITEA3 COMPACT, 2017-2020)

\bibliographystyle{IEEEbib}
\bibliography{bibliography}

\end{document}